# The twisting dynamics of large lattice mismatch van der Waals heterostructures

Mengzhou Liao[1,2*], Andrea Silva[3,9*], Luojun Du[1,7], Paolo Nicolini[2,10], Victor E. P. Claerbout[2], Denis Kramer[6], Rong Yang[8], Dongxia Shi[1,4,5], Tomas Polcar[2,3] and Guangyu Zhang[1,4,5*]

[1] *Beijing National Laboratory for Condensed Matter Physics and Institute of Physics, Chinese Academy of Sciences, Beijing 100190, China*

[2] *Department of Control Engineering, Faculty of Electrical Engineering, Czech Technical University in Prague, Technicka 2, 16627 Prague 6, Czech Republic*

[3] *National Centre for Advanced Tribology (nCATS), Department of Mechanical Engineering, University of Southampton, Highfield, SO17 1BJ, Southampton, United Kingdom*

[4] *Songshan Lake Materials Laboratory, Dongguan, Guangdong 523808, China*

[5] *School of Physical Sciences, University of Chinese Academy of Sciences, Beijing 100190, China*

[6] *Mechanical Engineering, Helmut Schmidt University, Hamburg, 22043, Germany*

[7] *Department of Electronics and Nanoengineering, Aalto University, Tietotie 3, FI-02150, Finland*

[8] *College of Semiconductors (College of Integrated Circuits), Hunan University, Changsha 410082, China*

[9] *CNR-IOM, Consiglio Nazionale delle Ricerche - Istituto Officina dei Materiali, c/o SISSA, 34136 Trieste, Italy*

[10] *Institute of Physics, Czech Academy of Sciences, Na Slovance 2, 182 21 Prague 8, Czech Republic*

\* Corresponding authors. E-mail: mengzlia@fel.cvut.cz; a.silva@soton.ac.uk; gyzhang@iphy.ac.cn

## Abstract

**Van der Waals (vdW) homo-/heterostructures are ideal systems for studying interfacial tribological properties such as structural superlubricity. Previous**

**studies concentrated on the mechanism of translational motion in vdW interfaces. However, detailed mechanisms and general properties of the rotational motion are barely explored. Here, we combine experiments and simulations to reveal the twisting dynamics of the MoS$_2$/graphite heterostructure. Unlike the translational friction falling into the superlubricity regime with no twist angle dependence, the dynamic rotational resistances highly depend on twist angles. Our results show that the periodic rotational resistance force originates from structural potential energy changes during the twisting. The structural potential energy of MoS$_2$/graphite heterostructure increases monotonically from 0° to 30° twist angles, and the estimated relative energy barrier is $(1.43 \pm 0.36) \times 10^{-3} \text{J}/\boldsymbol{m^2}$. The formation of Moiré superstructures in the graphene layer is the key to controlling the structural potential energy of the MoS$_2$/graphene heterostructure. Our results suggest that in twisting 2D heterostructures, even if the interface sliding friction is negligible, the evolving potential energy change results in a non-vanishing rotational resistance force. The structural change of the heterostructure can be an additional pathway for energy dissipation in the rotational motion, further enhancing the rotational friction force.**

## Introduction

Reducing parasitic energy consumption is highly desired with the increasing energy demand. Thus, reaching the "zero friction" world is crucial as friction-related processes consume almost one-third of the total energy in our world[1]. It is reported that realizing the incommensurate interface is the key to reaching the structural lubricity with vanishing interface friction[2-12]. 2D materials are unique for realizing incommensurate interfaces because of their atomic flat interface and the weak interlayer Van der Waals interaction[8-11, 13-14]. With naturally incommensurate interface structure, 2D heterostructures show even better superlubic performances[10-11]. Recently, twist angle independent near-zero interface friction has been found in large lattice mismatch 2D heterostructures. However, even if the interface friction is negligible, the total friction force for sliding is still nonzero as the edge pining effects in those heterostructures will dominate the friction process[11]. Those results indicate that more mechanisms should be revealed in the "post-structural lubricity" friction system to realize the actual zero friction state.

Previous studies concentrated more on the mechanism of translational motion in vdW interfaces. However, detailed mechanisms and general properties of the rotational motion are not well understood. Here, we combined experiments and simulations to characterize the twisting dynamic of the $MoS_2$/graphite heterostructure. In experiments, the atomic force microscope (AFM) was also used to directly rotate the $MoS_2$ domains on graphite to dynamically measure the rotational resistance force. We observed that the rotational resistance force is not constant, which has valleys with a 60° period and is rotationally asymmetric. Dynamic rotation experiments indicate that the rotational resistance force of 2D heterostructures can be modified by intrinsic torque arising from the symmetry of those two lattices. Thermally-induced rotation of the $MoS_2$/graphite heterostructure shows that heterostructures with initial twist angles between 1.5° and 18° will rotate towards around 1.5° after annealing, and those with initial twist angles >18° will not. Those results prove that the structural potential energy profile of twisted $MoS_2$/graphite heterostructure should have two plateaus around 30° to 18° and 0° to 1.5° twist angles, and it decreases rapidly from 18° to 1.5° twist angle. From experiment results, we can estimate the structural potential energy difference in the $MoS_2$/graphite heterostructure for twist angles between 0° and 30°, which is $(1.43 \pm 0.36) \times 10^{-3} J/m^2$. The large-scale molecular dynamics calculation results agree with the experimental observation and point out that the formation of Moiré structures on the graphene layer is the critical mechanism underpinning the energy balance of $MoS_2$/graphene heterostructures. Our results show that in twisting 2D heterostructures, the structural potential energy can modify the rotational resistance force even if the interface translation friction is negligible. The energy will not be conservative with the structural changing during the twisting, which can provide an extra pathway for energy dissipation.

**Growth and characterization of $MoS_2$/graphite heterostructure.**

The $MoS_2$/graphite heterostructures were prepared by an epitaxial growth technique described in our previous works[15-16] (see the Methods section). Fig. 1**a** shows an optical microscope image of the as-grown $MoS_2$/graphite heterostructure surface.

The shape of most MoS$_2$ domains on the graphite surface is aligned triangles. As shown in Fig. 1**a** and **e**, the step edge of graphite can stop the domain growth, which forms MoS$_2$ strips (please refer to Fig. S1 and Supplementary Note 1 for more details). Fig. 1**b** shows a typical AFM topographic image of the MoS$_2$/graphite heterostructure. The height of monolayer MoS$_2$ on graphite is 0.83 nm, agreeing with the previously reported values[17]. We used selected area electron diffraction (SAED) to characterize the lattice alignment of our MoS$_2$/graphite heterostructure. As illustrated in Fig. 1**c**, the hexagonal diffraction spots of both MoS$_2$ and graphite have the same orientation, indicating a twist angle of either 0° or 60° between the as-grown MoS$_2$ and graphite substrate. Raman and photoluminescence (PL) spectra in Fig. 1**d** also demonstrate the high quality of the samples. Fig. 1**e** reports a 3D diagram of our MoS$_2$/graphite heterostructure.

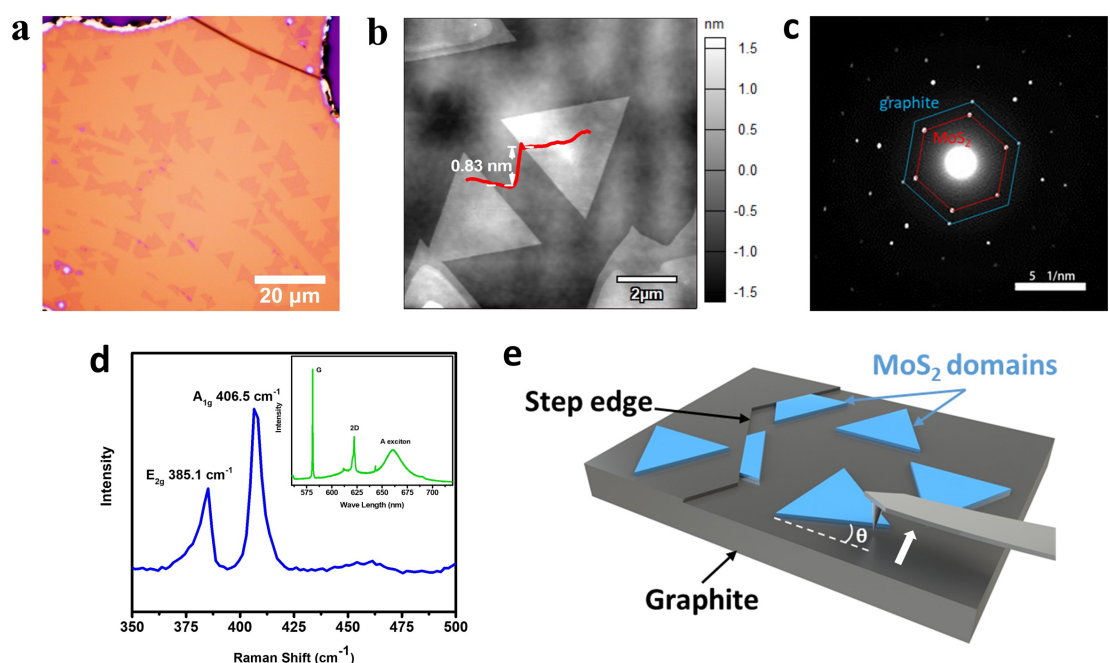

**Figure. 1 | Characterizations of MoS$_2$/graphite heterostructure. a** An optical microscope image of an as-grown MoS$_2$/graphite heterostructure sample. **b** An topography AFM image of the MoS$_2$/graphite heterostructure. **c** A SAED image of the MoS$_2$/graphite heterostructure. **d** Raman and PL spectra of the MoS$_2$/graphite heterostructure. **e** A 3D diagram of MoS$_2$/graphite heterostructure and the AFM manipulation method.

**Dynamic rotational resistance of MoS$_2$/graphite heterostructure.**

We used the AFM tip to apply the "push the edge" and the "drag the center" methods[11] to measure the rotational resistance of the MoS$_2$/graphite heterostructure dynamically in air. In the experiment, we found that due to the super low sliding friction, the translational energy barrier in the MoS$_2$/graphite heterostructure is smaller than the rotational energy barrier and that the MoS$_2$ domains favor rigid translation over twisting. Consequently, even pushing the edge on one strip side, MoS$_2$ domains will slide linearly instead of rotating in the inert gas environment (See the Video S1). However, the pure translation can be hindered by damaging the MoS$_2$ domain, using a fulcrum[18], or exposing the sample to air, and then the MoS$_2$ domains can be rotated. For more details, please refer to Supplementary Note 2. Fig. 2**a** shows the "push the edge" method: the tip moves perpendicular to the direction of the cantilever beam to push the edge on one side of the MoS$_2$ domain. In this method, we can determine the rotation center and angle by overlapping the AFM images from the same scan area before and after pushing. Fig. 2**b** shows the "drag the center" method: the tip first loads on the middle of one side of the MoS$_2$ domain and then move back and forth perpendicular to the direction of the cantilever beam to drag the MoS$_2$ domain. In the "drag the center" method, we usually created minor damage on the MoS$_2$ domain to make it rotate around it (see Video S1). The method to obtain the relation between the rotational resistance and the twist angles from raw AFM data is reported in Supplementary Note 4.

Unlike the sliding friction of MoS$_2$/graphite heterostructures with different twist angles[11], the dynamic rotational resistance force of the heterostructure is not constant. As shown in Fig. 2**c**, when rotating a MoS$_2$ domain on graphite anticlockwise by pushing the edge of it from -28.1° to 53.8° twist angles, we observed rotational resistance force valleys appear at twist angles around -11.5° and 50.2°. As shown in Fig. 2**d**, when continuously rotating the MoS$_2$/graphite heterostructure clockwise and anticlockwise by dragging the side middle of a MoS$_2$ domain, we also observed that the rotational resistance force is rotationally asymmetrical. Rotational resistance force valleys appear at different twist angles: around -9° for anticlockwise and 9° for clockwise. We also obtained similar results in MoS$_2$/h-BN heterostructure (see Fig. S6 in Supplementary Note 5), showing that the nonconstant dynamic rotational resistance is a general property of 2D heterostructures.

The periodic and asymmetrical rotational resistance force of the MoS$_2$/graphite heterostructure indicates that: structural potential energy change can modify the friction force offered by mechanisms such as the edge pinging effect in the rotational motion. The total rotational resistance force of the MoS$_2$/graphite heterostructure is the sum of the friction force and the force from intrinsic torque caused by the structural potential energy gradient.

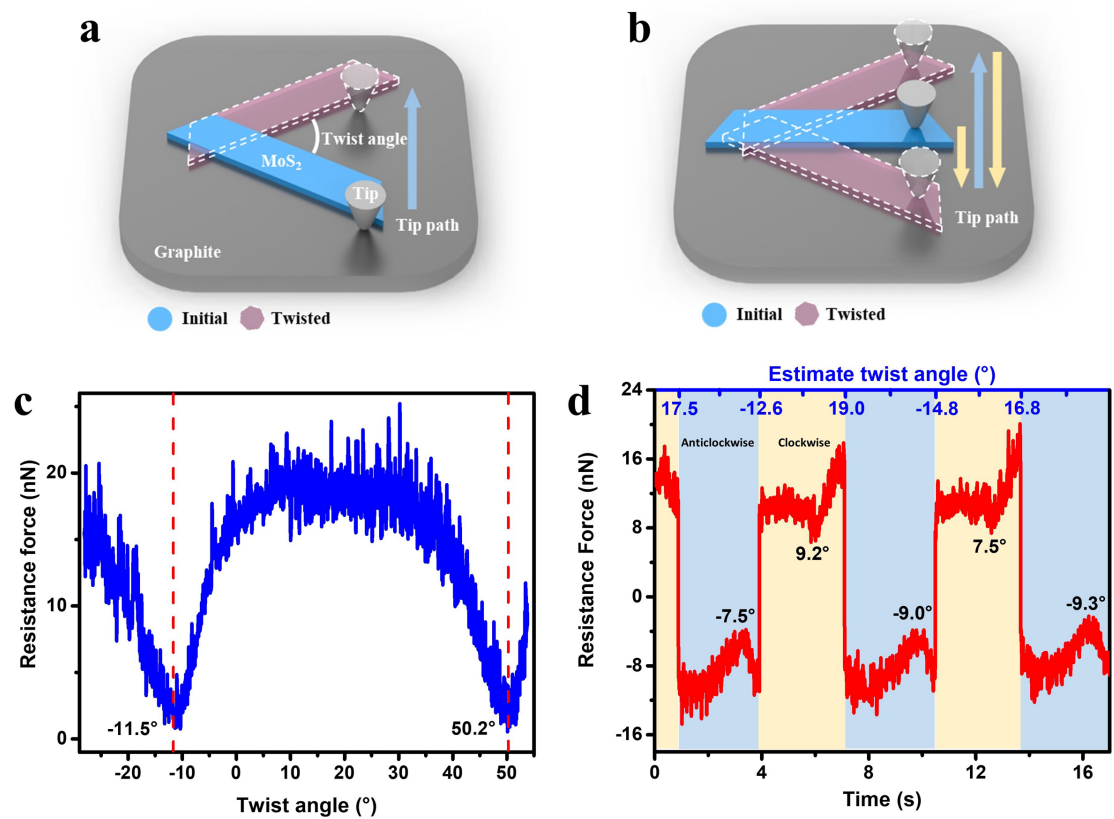

**Figure. 2 | Dynamic rotational resistance force measurements of the MoS$_2$/graphite heterostructure. a** and **b** Diagrams of the "push the edge" method (a) and the "drag the center" method (b). **c** Rotational resistance force of the MoS$_2$/graphite heterostructure as a function of the twist angle, measured by the "push the edge" method. Insert is an image that overlaps AFM images before and after rotation. **d** Clockwise and anticlockwise rotational resistance force of the MoS$_2$/graphite heterostructure, measured by moving continuously in the clockwise and anticlockwise directions using the "drag the center" method.

**Thermally-induced rotation of MoS$_2$ domains on graphite.**

In order to experimentally characterize the structural potential energy profile of twisted MoS$_2$/graphite heterostructures, we performed the thermally induced rotation on the heterostructures[19-20]. As shown in Fig. 3**a**, we first used AFM tips to rotate MoS$_2$ domains to random twist angles in air, and then we annealed them in a vacuum chamber at 450℃ for 4 hours to activate the thermally induced rotation towards the equilibrium configuration. The twist angles of MoS$_2$/graphite heterostructures are measured by comparing them with unrotated MoS$_2$ domains in AFM images. For more details of the thermally induced rotation experiment, please refer to Supplementary Note 3.

As shown in Fig. 3**b**, after annealing, we found some MoS$_2$ domains rotated close to 0°, while others kept their twist angles. In Fig. 3**c**, we show the statistic of twist angles for MoS$_2$/graphite heterostructures before and after annealing. We found that those heterostructures with initial twist angles below 18° will rotate close to 1.5°, and those with initial twist angles >18° will not rotate. This phenomenon indicates that: first, the total structural potential energy of 0° twist angle MoS$_2$/graphite heterostructure is the lowest; second, the total structural potential energy changes more rapidly for twist angles between 1.5° and 18° than for angles <1.5° or >18°, i.e., the intrinsic torque is the largest for the range between 1.5° and 18°. These results suggest that the structural potential energy profile of twisted MoS$_2$/graphite heterostructure should have two plateaus around 30° to 18° and 0° to 1.5° twist angles, which are connected by a sharp decrease from 18° to 1.5° twist angles.

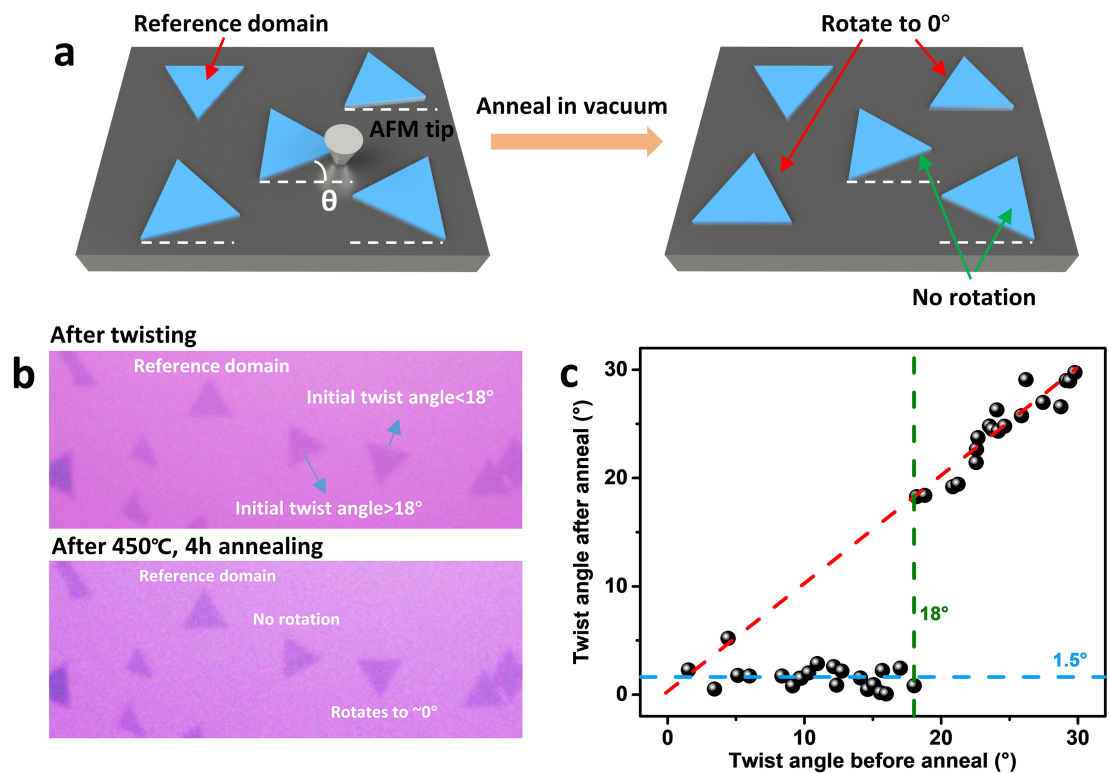

**Figure. 3 | Thermally-induced rotation in MoS$_2$/graphite heterostructure. a** A 3D diagram describing the experiment of thermally induced rotation. **b** Optical microscope images of MoS$_2$/graphite heterostructures after twisting (up) and annealing in a vacuum chamber (down). **c** Distribution of the twist angles of 41 MoS$_2$/graphite heterostructures before and after annealing. All twist angles are mapped into the [0,30°] interval, as the system periodicity is 60° and is symmetric around 0, i.e., the energy cannot depend on the sign of the angle.

Then we can explain the valleys in the rotational resistance force we measured in Fig.2**c-d**. When MoS$_2$ domains on graphite rotate from -30° to 30° (which we define as anticlockwise), intrinsic torque between twist angles from -18° to -1.5° will assist the rotation. Thus, the rotational resistance force will reduce, leading to the valley at around -11.5°. However, the intrinsic torque is rotationally asymmetrical because the structural potential energy releasing/absorption in the anticlockwise rotation will reverse at clockwise rotation, leading to the twisting asymmetrical rotational resistance force valleys in Fig. 2**d**. We also used the rotational resistance force from the "push the edge" method in Fig.2**c** to estimate the structural potential energy difference for twist angles between 0° and 30° MoS$_2$/graphite heterostructures (twisting barrier). The difference is $(1.43 \pm 0.36) \times 10^{-3} \text{J}/m^2$ . Like the "stick-slip" phenomenon in the translational motion, the twist angle controlled periodical structural potential energy may offer an extra energy dissipation pathway when twisting the heterostructure. For details of the barrier estimation from raw AFM data, please refer to Supplementary Note 6.

**Simulation results of the twisted MoS$_2$/graphene heterostructure.**

To understand the atomistic origin of the rotational energy landscape observed in the experiments, we performed constrained molecular mechanics simulations to control the evolution of the degrees of freedom of the heterostructure. The size of flakes in experiments is in the orders of microns, far beyond any atomistic methods capabilities. Hence in the simulations, we adopt periodic boundary conditions and focus on the interface contribution to the energy landscape. The challenge of simulating twisted vdW heterostructures lies in the large periodicity of the Moiré superstructure and the sensitivity of the total energy to the strain induced by the lattice mismatch between the pristine layers. The size of these systems places them beyond the capabilities of quantum mechanical methods, like density functional theory. To overcome these problems, we deploy a protocol developed and benchmarked in our previous studies[11,21], which has been shown to capture the relevant physics of this heterostructure system. In short, a classical force field refined to describe the interactions between graphene and MoS$_2$ is used to compute the energy of large-scale geometries, which contain from ~35k to 280k atoms. The large size of our supercells allows us to simulate almost any twist angle of MoS$_2$/graphene heterostructure with negligible influence due to the strain induced by the lattice mismatch between the pristine layers. For the detail of the simulation method, please refer to our previous work[21] and to Supplementary Note 7. Note that the simulated system has no edge due to the periodic boundary conditions adopted in the simulations. The edge effect in the twisting dynamics of the system considered is assessed in Supplementary Note 8. We parameterize a rigid model that efficiently describes the effect of unsaturated Moiré at the edges[22-25] and study the relative importance of elasticity and edge contributions to the rotational energy landscape as a function of flake size. Our analysis shows that while the edge effect is dominant at smaller sizes, the elastic contribution to the rotational energy landscape is dominant at the experimental size, supporting the analysis based on elasticity reported here; see Supplementary Note 8 for details.

Fig. 4**a** shows the simulated energy per unit area of the $MoS_2$/graphene heterostructure as a function of the twist angle. We can see from the results that: first, the energy of the twisted $MoS_2$/graphene heterostructure is monotonically decreasing from ±30° to 0°; second, the energy slope is larger between twist angles from -18° to -1.5° (1.5° to 18°), see dashed lines in Fig. 4**a**, which yields a relatively large intrinsic torque in this range. The simulation results fit the experiment well, except the energy barrier from simulations is ten times smaller than the experimental estimation. This quantitative difference originates from the model assumptions, as detailed in our previous benchmark[11, 21]. In Fig. 4**b**, we calculated the intrinsic torque as the first derivative of the energy (from the fitted curve in Fig. 4**a**). We define it as the intrinsic torque when rotating the $MoS_2$/graphene heterostructure anticlockwise. We found that the energy gradient reaches the negative peak in the region from -10° to -5°, which means that the intrinsic torque maximally reduces the anticlockwise rotational resistance force at those twist angles. We also noticed that the shape of Fig. 2**c** is not precisely the same as Fig. 4**b**. The reason for the difference, on the one hand, because if the frictional resistance force of the $MoS_2$ domain is smaller than the force offered by the intrinsic torque (yellow dash line in Fig. 4**b**, for example) during the rotation, then the $MoS_2$ domain will start to rotate spontaneously (i.e., a "rotational slip"). This self-induced rotation will drag the tip, suddenly decreasing the detected force, and finally leading to sharp force valleys as in Fig. 2**c**. On the other hand, the efficiency of the energy transition is not 100% (we will explain it later.), which means some other mechanisms can dissipate the energy. We also found that, in Fig. 4**b**, if we set the spontaneous rotation threshold to $1.95 \times 10^{-6} J/m^{2°}$, then intersections are 18° and 1.3°, which also fits the experiment quite well.

In light of this qualitative agreement between the computational energy landscape and the experimental observations, we analyzed the structures of all simulated twisted MoS$_2$/graphene heterostructures to reveal the underlying mechanism. In Fig. 4**c**, we calculated the root mean square displacements (RMSD) and standard deviation (STD) of all layers of atoms in twisted MoS$_2$/graphene heterostructures by using the isolated monolayers as references. The result shows that: first, the deformation of the MoS$_2$ layer is not strongly affected by the twist angle, which means the twist configurations in the heterostructure barely internal influence the structure of the MoS$_2$ layer; second, both RMSD and STD of the graphene layer in the heterostructure monotonously increase from twist angles ± 30° to 0°. In Fig. S8**a-c**, out-of-plane deformations mainly contribute to RMSD and STD in the graphene layer, which is thus the leading actor in shaping the energy landscape, which is supported by the bending modulus of graphene is much smaller than MoS$_2$, owing to its finite-thickness layer structures[26]. Fig. 4**d-f** shows the 3D diagrams of the deformation of graphene layers in the heterostructures at different twist angles. Note that the deformation of MoS$_2$ layers is amplified 20 times. From Fig. 4**d-f**, we can see that near 0° twist angle, the graphene layer in the heterostructure hosts a prominent a Moiré superstructure, which decays when the MoS$_2$/graphene heterostructure is twisted toward ± 30°. The Moiré structure also influences the interlayer distance, as shown in Fig. S8**d**. The formation of the long-wavelength Moiré structure at the graphene layer near 0° twist angle significantly releases the strain due to lattice mismatch while enhancing the coupling with the MoS$_2$ layer[11] so that the total energy is reduced. The graphene rigidity suppresses the short-wavelength Moiré structure at a 30° twist angle, in which the interlayer energy gain mentioned above is lost, and no stress is released. As a result, the total energy is the lowest at 0°, it sharply rises between 1.5° and 18°, and it reaches a plateau between 18° and 30°. If this driving force becomes smaller than a threshold determined by the experimental timescale (e.g., orange line in Fig. 4**b)**, observing the spontaneous rotation (e.g., during annealing) becomes negligible. This threshold force also explains why structures relax to nonzero angles.

The structural change in the heterostructure during the twisting must be carefully noticed because it can be a pathway of energy dissipation. Ideally, energy will be stored in the structure when twisting the heterostructure from twist angles with low structural potential energy to twist angles with high structural potential energy, and it will be released to assist the rotation when reversing the twisting direction. However, the efficiency of this energy transition is not 100%. When twisting the heterostructure, continuous changes of both the Moiré superstructure and the interlayer distance will introduce in- and out-plane vibrations, which will transfer part of the energy to thermal. Thus, in the rotational motion of 2D heterostructures, even if the interface is incommensurate with structural lubricity, the potential energy change can directly modify the friction force (for example, from the edge pinning effect) to form periodic rotational resistance force. Nevertheless, the structural change of the heterostructure can be an extra pathway for energy dissipation in the rotational motion, which will provide extra rotational friction force.

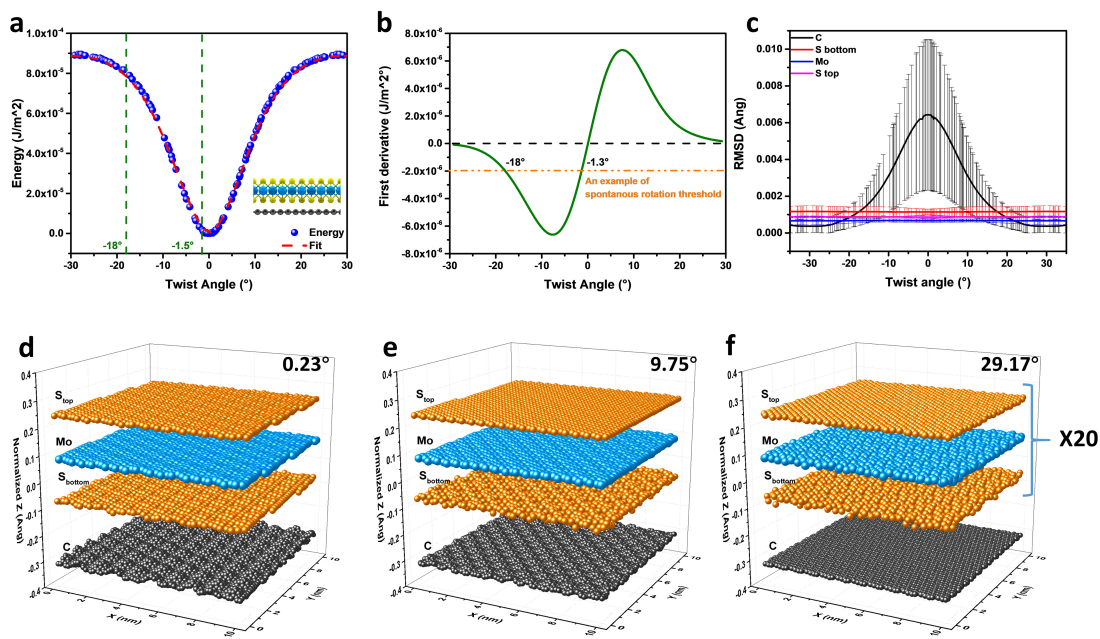

**Figure. 4 | Simulation results of the twisted MoS$_2$/graphene heterostructure. a** Normalized total energy per unit area of the MoS$_2$/graphene heterostructure as a function of the twist angle. The red dash line is the fitting, and the green dash lines mark the region between -1.5° and -18°. The inset shows a sketch of the heterostructure in the simulation. **b** The intrinsic torque as the first derivative of the fitted curve in **a**, the orange dash line is the assumed spontaneous rotation threshold. **c** Root mean square displacements of all layers of atoms in twisted MoS$_2$/graphene heterostructures as a function of twist angles. Error bars are the standard deviation. **d-f** 3D diagrams of all layers in the MoS$_2$/graphene heterostructure with different twist angles.

## Conclusions

In conclusion, the twisting dynamic of MoS$_2$/graphite heterostructures shows the complex origin of friction. Instead of just focusing on the interface corrugation to reach the near-zero potential energy surface[7, 27], changes in the structure of different twist configurations also need to be considered to reach near-zero friction in the rotational motion between atomic flat crystalline surfaces. On the one hand, the structural potential energy variation in the dynamic twisting can directly modify the rotational friction; on the other hand, the structural change of the material in the twisting can provide a pathway of energy dissipation, which can be the primary origin of the rotational friction. Thus, The structural effect via twisting needs to be avoided to reach zero friction. The atomic flat amorphous/amorphous or amorphous/crystalline interfaces could be promising candidates to reach simultaneous near-zero translational and rotational motion friction.

**Methods**

*Sample preparation*. MoS$_2$ domains were grown by three-temperature-zone chemical vapor deposition. S (Alfa Aesar, 99.9%, 4 g) and MoO$_3$ (Alfa Aesar, 99.999%, 50 mg) powders were used as sources, loaded separately in two inner tubes and placed at zone-I and Zone-II, respectively. Substrates were loaded in zone-III. During the growth, Ar/O$_2$ (gas flow rate: 75/3 sccm) was flowed as carrying gases and temperatures for the S-source, MoO$_3$-source and wafer substrate were 115 °C, 530 °C, and 930 °C, respectively. The graphite substrate was mechanically exfoliated from HOPG or Graphenium graphite (*Manchester Nanomaterials*) in this experiment.

*Sample characterizations*. AFM measurements were performed on Asylum Research Cypher S with AC240TS-R3 tips. We used standard Sader's method to calibrate the cantilever in the vertical direction and a non-contact method for the lateral direction[28-29]. The velocity of the tips was 0.6μm/s. PL and Raman characterizations were performed in a Horiba Jobin Yvon LabRAM HR-Evolution Raman system. The laser wavelength was 532 nm. SAED was performed in a TEM (Philips CM200) operating at 200 kV.

*Molecular mechanics simulations*. All energy minimizations of the rotated heterostructures have been performed using molecular dynamics by means of the LAMMPS package[30] using the conjugate gradient algorithm, where the energy tolerance was set to $1 \times 10^{-15}$. The reactive empirical bond order (REBO) potential was used for graphene[31], whereas the Stillinger-Weber potential was used for $MoS_2$[32]. To model the vdW interactions, we used a modified interlayer Lennard-Jones potential parameterized in Ref.[21]. The geometries are described in detail in the Supplementary Note 7.

Supporting Information

The evidence of $MoS_2$ domains' growth stopped by graphite step edges. Details of the method used to rotate the MoS2 domains on graphite, thermally induced rotation, rotational resistance calibration, calculation of the structural potential energy difference, molecular mechanics simulations. Rotational behavior of the $MoS_2$/h-BN heterostructure. Discussion of the Edge effects.

**Acknowledgments**

GZ thanks for the support from the National Science Foundation of China (NSFC, Grant No. 11834017 & 61888102) and the Strategic Priority Research Program of CAS (Grant No. XDB30000000). ML thanks the support from ESI Fund, OPR DE International Mobility of Researchers MSCA-IF III at CTU in Prague (No: CZ.02.2.69/0.0/0.0/20_079/0017983). LD. gratefully acknowledges the financial support by the Academy of Finland (Grant No. 333099). DS thanks for the support from NSFC (Grant No. 61734001). ML, PN and TP acknowledge support from the project Novel nanostructures for engineering applications CZ.02.1.01/0.0/0.0/16_026/0008396. This work was supported by the Ministry of Education, Youth and Sports of the Czech Republic through the e-INFRA CZ (ID:90140). TP, VEPC and AS acknowledge support from the European Union's Horizon2020 research and innovation program under grant agreement No. 721642: SOLUTION. RY thanks the support from NSFC (Grant No. 12074412 and 62122084). The authors acknowledge the use of the IRIDIS High Performance Computing Facility, and associated support services at the University of Southampton, in the completion of this work.

**Author contributions**

Guangyu Zhang and Tomas Polcar supervised the research; Mengzhou Liao performed the AFM measurements and data analysis; Luojun Du performed the sample growth,

TEM and spectroscopic characterizations; Andrea Silva, Paolo Nicolini, and Victor E. P. Claerbout performed the simulations; Mengzhou Liao and Andrea Silva wrote the manuscript; Rong Yang and Dongxia Shi helped in lab management; All authors commented on the manuscript.

## Competing interests

The authors declare no competing interests.